\documentclass[11pt]{article}
\usepackage{moriond,epsfig}

\def\Journal#1#2#3#4{{#1} {\bf #2}, #3 (#4)}

\def\PRD{{\em Phys. Rev.} D}

\def\be{\begin{equation}}
\def\ee{\end{equation}}
\def\bea{\begin{eqnarray}}
\def\eea{\end{eqnarray}}

\begin{document}
\vspace*{4cm}
\title{SINGLE TOP PRODUCTION AT $\sqrt s=7$~TeV}

\author{REBECA GONZALEZ SUAREZ, on behalf of the CMS and ATLAS collaborations }

\address{IIHE - VUB,  Pleinlaan 2 \\
1050 Brussels, Belgium}

\maketitle\abstracts{
The production of single top quarks occurs via three processes: t-channel, s-channel and tW associated production. The LHC experiments have observed single top production via t-channel at 7 TeV and measured its cross section, providing a measurement of $|V_{tb}|$ with an uncertainty at the 10\% level. Studies are in place to observe tW associated production with a sensitivity close to 3$\sigma$ and the first limits on the production cross section for s-channel are set. Other studies based on single top topologies, like flavor changing neutral currents (FCNC) are also being performed.
}

\section{Introduction}
In hadronic colliders, top quarks are mostly produced in pairs via strong interaction. An alternative production via the weak interaction, that involves a Wtb vertex, leading to a single top quark final state, proceeds by three different mechanisms: t-channel, associated production with a W boson (tW), and s-channel.
The study of single top quark processes allows for a measurement of the CKM matrix element $|V_{tb}|$ without assumptions about the number of quark generations,  single top processes are also sensitive to many models of new physics.
Single top production was first observed at the Tevatron in 2009, in a combination of t and s-channel; the LHC experiments, ATLAS~\cite{atlas} and CMS~\cite{cms}, have observed single top in the t-channel and measure its production cross section. Studies are also in place to study of tW associated production and s-channel.

\section{t-channel}
The t-channel production is the process  with the highest cross section at the LHC, $\sigma_t = 64.6^{+3.3}_{-2.6}$pb~\cite{kidonakist}. This process is studied in final states with lepton+jets  signature. The signal is characterized by one isolated muon or electron and missing transverse energy ($E_{T}^{miss}$), plus a central jet coming from the decay of a b quark and an additional light jet from the hard scattering process, that is often forward. A second b-jet produced in association to the top quark can be present as well, with a softer $p_{T}$ spectrum with respect to the one coming from top decay.

\subsection{Selection criteria}
To define the signal region, events with exactly 1 isolated lepton ($e$,$\mu$) and 2 or 3 jets, one of them identified as coming from the decay of a b-quark (b-tagged) are selected. ATLAS~\cite{ATLAS101}  applies a cut on the missing $E_{T}$ of the event, $E_{T}^{miss} > 25$~GeV; while CMS~\cite{TOP-11-021} selects events with $E_{T}^{miss} > 35$~GeV in final states with electrons and applies a cut on the transverse mass of the W boson, $m_T(W) > 40$~GeV, in the final states with muons. Then ATLAS applies a triangular cut:  $m_T (W) >$(60~GeV - $E_{T}^{miss})$, while CMS uses the invariant mass of the reconstructed top quark, $130 < m_t <  220$~GeV. Other jet and b-tagging multiplicities are used in background estimations and as control regions. The main backgrounds that contribute to the analysis are W boson production in association with jets (W+jets), top pair ($t\bar{t}$) production and multijets (QCD) events. Background from $t\bar{t}$ and other processes like Z+jets, other single-top processes, and di-boson production, are estimated from simulation and normalized to their theoretical cross sections, while dedicated methods are applied to estimate the W+jets and QCD contributions.

\subsection{Background estimation and Signal extraction}
The QCD contribution is estimated via a maximum likelihood fit to the $E_{T}^{miss}$ ($e,\mu$ ATLAS, $e$ CMS) or $m_T(W)$ distribution ($\mu$ CMS). 
The template for QCD is obtained in data by inverting the isolation on muons and either requiring the electron to fail some of the quality requirements (CMS), or replacing it by a jet passing similar requirements (jet-electron model, ATLAS). For all other processes ($t\bar{t}$, W/Z+jets, di-bosons), the templates are obtained from simulation. \\
To estimate the W+jets background, ATLAS uses the distributions from simulation and extracts an overall normalization factor and the flavor composition from data. CMS extracts the W+jets shapes and the normalization from the events that fail the reconstructed top quark mass cut, subtracting other backgrounds.\\
For the signal extraction, ATLAS uses a set of discriminant variables as input to a cut based analysis and a Neural Network multivariate analysis. The main variables are the reconstructed top quark mass, the pseudorapidity of the light (untagged) jet, $|\eta_{jÕ}|$ and the transverse energy of the light (untagged) jet. CMS carries out the extraction in a different way, performing a maximum likelihood fit to the distribution of  $|\eta_{jÕ}|$.

\subsection{Results}
The dominant sources of systematic uncertainty arise from detector simulation and object modeling, where the higher effect comes from jet energy scale (9\%) b-tagging (18\% ATLAS, 3\% CMS), and jet energy resolution (]6\% ATLAS, 1\% CMS). Another main source of uncertainty are theoretical uncertainties. In the case of CMS the renormalization and factorization scale Q$^2$ (7\%) is the most important, while for ATLAS, the effect of initial and final state radiation (14\%) is the dominant one, followed by the choice of generator (11\%) and parton shower modeling (10\%).\\
The baseline result for ATLAS comes from the analysis based on cuts, that uses final states with 2 and 3 jets, and has a slightly smaller overall expected uncertainty. The value of the cross section measured by this analysis on 0.7fb$^{-1}$ of data is: $\sigma_t = 90^{+9}_{-9}(stat)^{+31}_{-20}(syst) =90^{+32}_{-22}$pb.\\
The cross section measured by CMS, using 1.14~fb$^{-1}$ in final states with muons and 1.5~fb$^{-1}$ in final states with electrons, is: $\sigma_t = 70.2\pm5.2(stat)\pm10.4(syst)\pm3.4(lumi)$pb, and from this value, CMS provides an estimation of  $|V_{tb}|$, assuming $|V_{ts}|,|V_{td}| << |V_{tb}|$:  $|V_{tb}| = \sqrt{\frac{\sigma_t}{\sigma_t^{th}}} = 1.04\pm0.09(exp.)\pm0.02(th.)$

\begin{figure}[h]
\begin{center} 
\psfig{figure=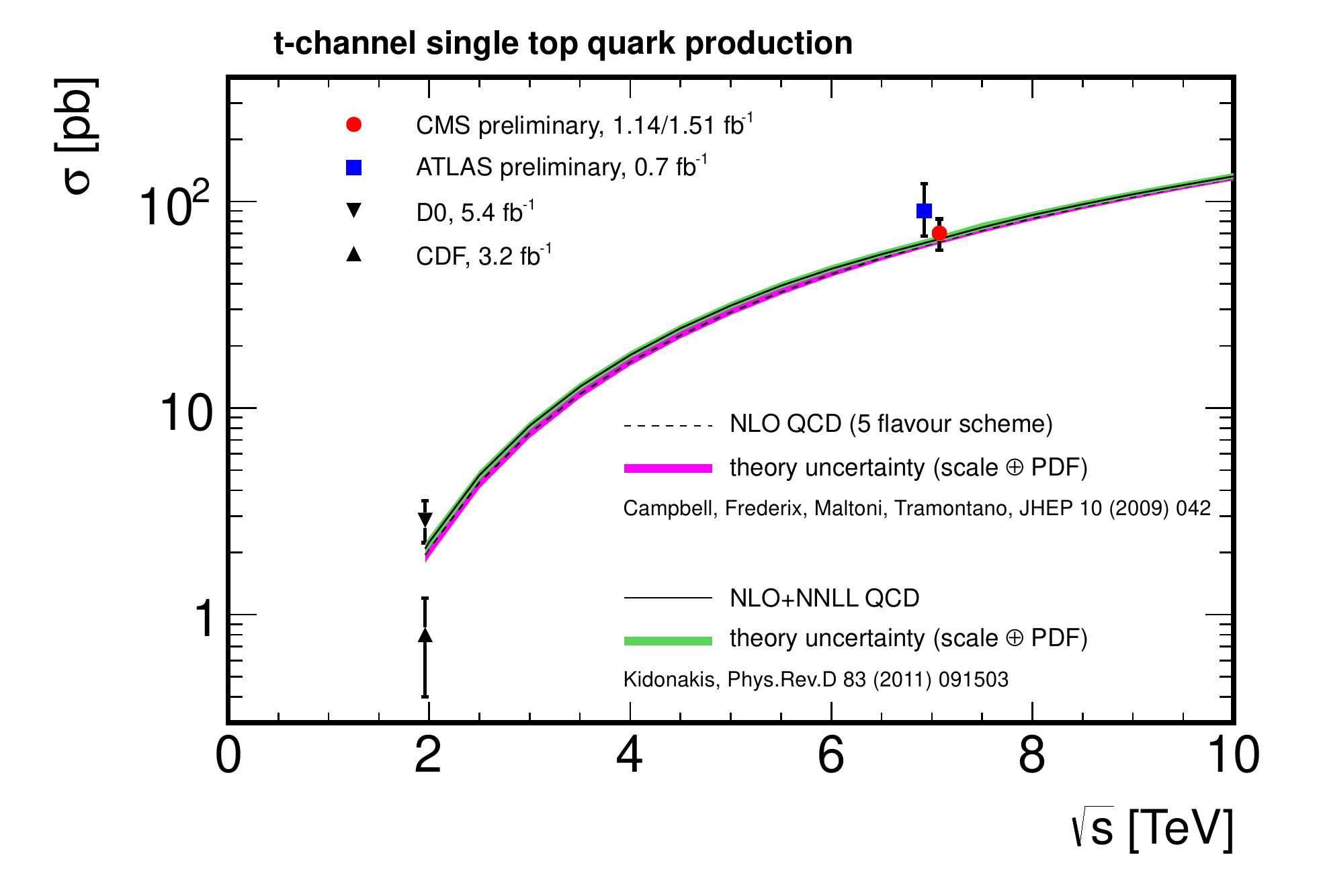, height=65mm} 
\caption{Single top t-channel cross section measurements from ATLAS and CMS, including CDF and D0 results.
\label{fig:xsect}}
\end{center} 
\end{figure}

\section{tW associated production}
The associated production of single top quarks with a W boson (tW), inaccessible at the Tevatron, is the second in terms of cross section at the LHC, with $\sigma_{tW} = 15.7^{+1.3}_{-1.4}$pb~\cite{kidonakis}. It has an interesting topology, background to $H\rightarrow WW $searches; and has never been observed. \\
The leptonic decays of the W bosons are studied at the LHC, in signatures with two leptons, $E_{T}^{miss}$ and one jet originating from the hadronization of a b-quark. The main backgrounds for this final state are $t\bar{t}$ production and Z+jets, with small contributions from di-bosons, other single top channels, W+jets and QCD.

\subsection{Event Selection and background estimation}
Events with two leptons, electrons or muons, and one jet are selected~\cite{ATLAS104,TOP-11-022}. The analysis performed by CMS requires the jet to be b-tagged. A substantial amount of $E_{T}^{miss}$ is expected to be present in the event, due to the presence of two neutrinos in the final state, therefore a cut on the $E_{T}^{miss}$ of the event is applied: $E_{T}^{miss}>$ 50 (30) ATLAS (CMS). To remove Z+jets background, events inside the Z mass window, $81 < m_{ll} < 101$~GeV, are rejected in the $ee$ and $\mu\mu$ final states. \\
To complete the definition of the signal region CMS applies cuts on two extra variables, the $p_{T}$ of the system formed by the leptons, the jet and  $E_{T}^{miss}$, and $H_{T}$, defined as the scalar sum of the $p_{T}$ of the leptons, the $p_{T}$ of the jet and $E_{T}^{miss}$. ATLAS has a dedicated anti $Z\rightarrow \tau\tau$ cut:  $\Delta\phi(l_1,E_{T}^{miss})+\Delta\phi(l_2,E_{T}^{miss}) > 2.5$.\\
CMS estimates the Z+jets background from data, using events in and out of the Z mass window, and uses two $t\bar{t}$ enriched control regions (events with two jets with either one or both of them b-tagged) that are considered in the significance calculation to constrain $t\bar{t}$ contamination and b-tagging efficiency.
ATLAS also estimates Z+jets background from data, using the so called ABCDEF method, where orthogonal cuts on 2 variables ($m_{ll}$ and $E_{T}^{miss}$) define signal and background enriched regions. The contribution from fake leptons coming from W+jets (single) and QCD (double-fake) is estimated in ATLAS using the matrix method ($< 1$\% effect). Finally, an estimation of the $Z\rightarrow \tau\tau$ background from data is performed and a scale factor for $t\bar{t}$ is obtained from events with two jets.

\subsection{Results}

The main sources of systematic uncertainty for CMS are the ones associated to the b-tagging (10\%) and Q$^2$ (10\%); while for ATLAS,  the jet energy scale (35\%) and resolution (32\%), and background normalization are dominant.\\
ATLAS set a 95\% CL limit on the production of tW of $\sigma_{tW} < 39.1(40.6)$pb obs.(exp.) with an observed significance of 1.2$\sigma$,  estimating a value of the cross section of  $\sigma_{tW} = 14^{+5.3}_{-5.1}(stat.)^{+9.7}_{-9.4}(syst.)$~pb using 0.7~fb$^{-1}$ of data.\\
CMS, with 2.1fb$^{-1}$ has an observed (expected) significance of 2.7$\sigma$ (1.8$\pm$0.9$\sigma$), with a measured value of the cross section and 68\% CL interval of $\sigma_{tW} = 22^{+9}_{-7} (stat+sys)$pb.

\section{s-channel}

The s-channel, $\sigma_{s} = 4.6\pm0.3$pb~\cite{kidonakiss} , is a process sensitive to several models of new physics, like $W^{'}$ bosons or charged Higgs bosons. Similar to tW production, it has not been observed yet. \\
It has a very challenging lepton+jets signature, difficult to separate from the backgrounds ($t\bar{t}$, W+jets and QCD). At the LHC, ATLAS performs an analysis~\cite{ATLAS118}  using similar objects and preselection as for the t-channel; as well as the same background estimations for QCD and W+jets. After the final selection, made with a set of cuts, the signal purity is 6\%. An upper limit on the production cross section is set at 95\% CL, $\sigma_t < 26.5 (20.5)$~pb observed (expected) with 0.7~fb$^{-1}$ of data.

\section{Other single top studies: FCNC single top quark production}

ATLAS performs a search for Flavor Changing Neutral Currents (FCNC)~\cite{FCNC} in the 1 jet bin, using leptonic decays. Events are classified using a neural network where the most significant variables are the $p_T$ of the W boson, $\Delta R_{(b-jet,lepton)}$ and lepton charge. Over an integrated luminosity of 2.05~fb$^{-1}$, no excess is observed over the Standard Model expectations and limits are set on the coupling constants $\kappa_{ugt}/\Lambda$ and  $\kappa_{cgt}/\Lambda$, and on the branching fractions $t\rightarrow ug$ and $t\rightarrow cg$: $\sigma(qg\rightarrow t)\cdot B(t\rightarrow Wb) < 3.9$pb (95\% CL),  $B(t \rightarrow ug) < 5.7\cdot10^{-5}$, $B(t \rightarrow cg) < 2.7\cdot10^{-4}$.

\section{Conclusion and outlook}
ATLAS and CMS have a broad program of single top physics: the cross section of the t-channel production has been measured, allowing to also measure $|V_{tb}|$ at the 10\% level; the first hints of tW associated production have been studied with a significance close to 3$\sigma$; the first upper limits on s-channel production have also been set; and finally, there are already results from other single top studies, the latest concerning the FCNC single top quark production.

\end{document}